\documentclass[12pt,a4paper]{article}

\usepackage{graphicx}
\usepackage{amsmath}
\usepackage{amsfonts}
\usepackage{amssymb}
\def\simleq{\; \raise0.3ex\hbox{$<$\kern-0.75em \raise-1.1ex\hbox{$\sim$}}\; }
\def\simgeq{\; \raise0.3ex\hbox{$>$\kern-0.75em \raise-1.1ex\hbox{$\sim$}}\; }
\newcommand{\eV}{{\rm eV}}
\newcommand{\keV}{{\rm keV}}
\newcommand{\MeV}{{\rm MeV}}
\newcommand{\GeV}{{\rm GeV}}
\newcommand{\TeV}{{\rm TeV}}
\newcommand{\EeV}{{\rm EeV}}
\newcommand{\erg}{{\rm ergs}}

\newcommand{\kpc}{{\rm kpc}}
\newcommand{\pc}{{\rm pc}}
\newcommand{\cm}{{\rm  cm}}
\newcommand{\km}{{\rm km}}
\newcommand{\muG}{\mu{\rm G}}

\newcommand{\s}{{\rm s}}
\newcommand{\yr}{{\rm yr}}
\newcommand{\sr}{{\rm sr}}

\begin{document}
\title{Sgr A East as a possible high energy neutron factory in the Galactic Centre}
\author{Dario Grasso$^{1,2~\dag}$,    Luca Maccione$^{2,3~\ddag}$}
\maketitle
\noindent
$^1${ Scuola Normale Superiore, P.zza Dei Cavalieri, 7,  I-56126 PISA}\\
$^2$ {I.N.F.N. Sezione di Pisa}\\
$^3$ Dip. di Fisica ``E.~Fermi", Universit\`a di Pisa, Largo B.~Pontecorvo, 3,  PISA
\vskip0.3cm \noindent
$^\dag$ d.grasso@sns.it; \ $^\ddag$ luca.maccione@pi.infn.it
\noindent
\begin{abstract} 
Sgr A East is a supernova remnant  located within few parsecs from the 
Galactic Centre (GC).  There are good reasons to believe that this object is the source
 of the $\gamma$-ray excess detected by HESS  in the direction of the GC meaning that  
Sgr A East is likely to be an efficient  Cosmic Ray (CR) accelerator.
Some observations suggest that  strong magnetic fields may be present in that 
region which may allow the acceleration of composite nuclei in Sgr A East beyond the EeV. 
We show that, if this is case, EeV neutrons  should be effectively produced by
the photo-disintegration of  Ultra High Energy nuclei onto the 
infrared photon background (with temperature  $\sim 40$ K) in which  Sgr A East is embedded. 
 Neutrons with such an energy  can reach the Earth before decaying
and may be detectable  under the form of a CR point-like excess in the direction of the GC.
 We determine the expected energy spectrum and the amplitude of this signal showing 
that it may be measurable by the AUGER observatory. 
\end{abstract}

\section{Introduction}

One of the main puzzles in cosmic ray physics concerns the origin of cosmic rays (CRs)
with energies between the knee ($E_{\rm knee} \sim 3 \times 10^{15}~\eV$) and the ankle 
  ($E_{\rm ankle} \sim 6 \times 10^{18}~\eV$) of the CR spectrum.
It is generally accepted that CRs with energy up to the knee are accelerated by  the `so-called'
Fermi first order  mechanism (for a review see e.g. \cite{Drury83}) 
in  the supersonic outflow of galactic  SuperNova Remnants (SNRs).  
This wisdom is motivated not only  by the well  known coincidence  between the  galactic  SN
 energy output   and that of galactic CRs but also by  the slope and the regularity of the CR 
 energy spectrum over several energy decades.   The observation of radio and
 $\gamma$-ray emissions from a number of SNRs provides further evidence of the validity
 of this scenario. It   is more uncertain, however,  what is the maximal  energy  $E_{\rm max}$
 at which particles can be accelerated  in this  kind of sources.  
 The continuity of the CR spectrum up to ankle  suggests that the origin of CR  above the knee
 is closely related to that below the knee meaning that  $E_{\rm max}$ may be larger than 
 the EeV.
  
 One of the  crucial quantities  which enters to determine  $E_{\rm max}$
  is the magnetic field strength in the region  where the  SNR shell is expanding through.
  It is known that, if the mean value of the galactic magnetic field ($B_{\rm gal}
 \simeq 5~\muG$) is adopted, the theoretically estimated $E_{\rm max}$ cannot exceed
 the   Lagage and Cesarsky bound  $E_{\rm max}^{\rm LC} \sim Z \times  10^{16}~\eV$ \cite{Lagage}.
Stronger  magnetic fields  may however be found in some dense regions of the Galaxy, like the Galactic Centre (GC) \cite{YZ96} or in dense molecular clouds \cite{Heiles05}.
  It is also possible that  non-linear effects,  e.g.  cosmic-ray streaming instabilities \cite{Bell:01},  
  give rise to an amplification of the magnetic field in the acceleration region allowing to
  reach CR energies exceeding the Lagage and Cesarsky bound.  Even in this case, however,
  the largest CR energies are expected to be reached  in SNRs placed in regions with the
   strongest  magnetic field backgrounds. 
 
 It would be extremely interesting to find an observational evidence showing that CR acceleration
 up to Ultra High Energies (UHEs)  ($E \simgeq 10^{18}~\eV$) takes place in some galactic SNR. 
  Unfortunately, even EeV protons are significantly deflected in the galactic magnetic field so that
 the angular correlation between the arrival direction of these particles and the position of their
  sources should be spoiled. A  possible way out may, however, be found if  EeV secondary 
  neutrons are efficiently produced  by the scattering of the UHE primary nuclei with the matter 
  and the radiation surrounding the SNR. 
 Neutrons with such an energy can travel over galactic distances without undergoing significant
 decay and  deflections.  Although high energy neutrons and protons give rise to indistinguishable showers in the atmosphere, a galactic neutron source may  be recognised by the extensive air shower
  experiments under the form of a localised excess of CRs with energy $\simgeq 1~\EeV$.    
 Clearly, a complete identification of the source requires a matching
 between the position of the CR excess and that of a radio, X-ray  and, possibly, $\gamma$-ray
 emitting  SNR \cite{Drury94}.       
 
 The SNRs which are  the most promising  sources of  EeV neutrons are those residing  in the
 densest regions of the Galaxy, where strong magnetic fields as well as  thick radiation
 and gas targets can presumably be found.  Active SNRs  in the proximity of the
  Galactic Center (GC)  and/or in the nearby of dense molecular clouds are  natural 
 candidates to the role of  EeV neutron factories. 
 
 In this respect, as it was  recently suggested in \cite{Crocker},  a particularly interesting 
 system is Sgr A East.  
Sgr A East is a SNR located within few parsecs from the dynamical centre of the Galaxy.
Detailed radio and X-ray observations \cite{BeppoSax,Chandra} allowed to conclude
that this object is a mixed-morphology (radio shell + X-ray emitting core) remnant of a
single type II SN explosion which took place between $10^{4}$ and $10^{5}$ years ago
(the light propagation time is subtracted).  
This SNR is embedded within a dense ionised gas halo and in a background of
Infrared (IR) radiation due to dust emission at the temperature $T \sim 40$ K \cite{Zylka95, Philipp99}.  Furthermore, the expanding Sgr A East radio shell is interacting  with a dense 
molecular cloud. Strong magnetic fields, as large as few milliGauss, have been observed in
that region \cite{YZ96}.  
 
 A $\gamma$-ray emission has been also recently observed by the HESS Cerenkov telescope 
 in the direction of the GC \cite{HESS}.
 Although the limited angular resolution reachable with this kind of 
 observations did not allow  a firm identification of the source of this emission,  several 
 arguments points to Sgr A East as the most plausible source.   The most convincing 
 among these arguments is  based on the  energetic of the emission observed by HESS.
 In Sec. \ref{gammas} we show, in fact,  that  the $\gamma$-ray flux measured by HESS 
 practically coincides  with that expected from the  decay of neutral pions produced in the
 hadronic  collisions of UHE protons which are  shock accelerated by a SNR with the same 
 characteristics of Sgr A East.
 On the basis of this  argument, as well as other arguments listed in Sec.\ref{gammas},
 we assume  that Sgr A East accelerates nuclei with 
 a power law spectrum with index equal to that of the $\gamma$-ray excess observed by HESS.  
  
 In Sec.\ref{Emax} we argue that the spectrum of CRs accelerated in Sgr A East
 may extend beyond the EeV.
 
EeV neutrons may be produced as secondary particles of UHE nuclei
  in Sgr A East by two main processes: the Photo-Disintegration (PD)   of
composite nuclei onto the 40 K IR photon background present in that region and the $pp$
inelastic scattering of UHE protons onto the dense hydrogen gas. 
In this paper (see Sec.s \ref{PD} and \ref{neu_pp}) 
we estimate the expected neutron flux reaching the Earth  produced by  both processes.
We show  that PD  should provide  the dominant, if not the unique, contribution. 
This contribution was not computed in a previous work on the subject \cite{Crocker}.
   
 In Sec.\ref{AUGER} we estimate the  event  rate    
 expected at  the Pierre AUGER Observatory \cite{Auger} due to neutrons from Sgr A East.

The possibility that EeV neutrons could be produced by a SNR in the GC giving rise to a localised 
CR excesses around the EeV has been already suggested in several papers (see e.g. 
\cite{Hayashida:1998qb,Bednarek,Anchordoqui}).  In particular, Crocker   {\it et al.} \cite{Crocker}
first proposed Sgr A East as a possible high energy $\gamma$-ray and EeV neutron source,
although the contribution of nuclei PD to the neutron flux was not determined in 
their paper.
All these works, however, tried to interpret  the EeV  CR anisotropy claimed by the AGASA
 \cite{Hayashida:1998qb} and SUGAR \cite{Bellido:2000tr} experiments. 
  This is not what we do in this paper. Actually, in Sec.\ref{discussion} we argue that
both  AGASA and  SUGAR anisotropies  can hardly be ascribed to a neutron emission
 from the Galactic centre.  In particular we show that the EeV neutron flux estimated  in \cite{Crocker} is likely to be incompatible with the energetic of Sgr A East as  constrained  by Chandra \cite{Chandra} and HESS \cite{HESS} observations. 
Here we only use HESS observations to normalise the primary spectrum of nuclei
accelerated in Sgr A East assuming that this spectrum remain unchanged up to UHEs.  
Although our prediction for the EeV neutron flux is much lower than than estimated
in \cite{Crocker} we show that such a flux should however be detectable by AUGER.
In Sec.\ref{conclusions} we summarise our conclusions.
 
\section{Physical conditions in the  Sgr A East region}\label{observations}

 Sgr A East has been observed at several wavelengths.
 Its distance has been determined to be $d \simeq 8~\kpc$ \cite{Reid93}.

In the radio  this object appears in the form of  a non-thermal shell of ellipsoidal shape.
The major axis of the ellipse  has a  length of about 10 parsecs and runs almost parallel to the 
galactic plane.
 The East side of the shell is expanding through  the dense molecular cloud M-0.02-0.07 . The 
 compressed dust/molecular ridge produced by the interaction of these objects give rise to a 
 OH maser emission . Yusuf-Zadeh et al.  \cite{YZ96} found some evidence of 
 Zeeman splitting in the spectrum of this emission and inferred a magnetic field strength of 2-4 mG.   Even if such a high field strength  is not representative of the entire cloud,  the  
 synchrotron loss time of relativistic electrons must be considerably smaller than Sgr A East
 age implying that particle acceleration is still active in that region. 
 
 Detailed Far InfraRed (FIR) and Medium InfraRed (MIR) observations of the GC region
 have also been performed allowing to identify several diffuse and point-like sources.  
  Here we are mainly interested in the diffuse radiation. 
 In the Nuclear Bulge ($R \simleq 300~\pc$) the IR background  appears to be dominated by cold dust  emission with temperature $T \sim 20 - 50~$K  (FIR).   This is the true 
 black-body emission as, in the few central parsecs of the Galaxy, and especially  in the 
 dense Sgr A East dust ridge, the optical depth is much smaller than unity so that the
 intense UV radiation from the hot stars present in that region is effectively
 thermalised.  
 Warm dust  emission with temperature up $\sim 200~$K dominates  only in the 
central parsec of the Galaxy.  According to the work of Davidson  \cite{Davidson92}
$T \sim 40~$K for 
$R \sim 8~\pc$ and $T \sim 50~$K for $R \sim 1~\pc$ (see also \cite{Zylka95, Philipp99}).
 Therefore 40 K  must be the  temperature in the NE  side of Sgr A East shell.
 This is also the temperature which, according to Gordon at al. \cite{Gordon93},  is typical for the molecular clouds in the Nuclear Bulge. 
  
The spectacular  Chandra \cite{Chandra} X-ray observations, in the energy range $2-10~\keV$, 
with an angular resolution  better than  1'',   confirmed that Sgr A East is the remnant of a
single  type II SN with age between $10^{4}$ and $10^5$ years. 
The  total luminosity  ($\sim 8 \times 10^{34}~\erg ~\s^{-1}$ in the Chandra energy range) and  
the spectrum of this source allowed to establish that its progenitor was a main sequence star of mass of $13-20~M_\odot$ with a metallicity  four times  larger than  that of the Sun. 
The energy of the X-ray emitting plasma was estimated to be  $10^{49}$ ergs, which is
compatible with a total kinetic energy released in  the explosion  of  $10^{51}$ ergs,  as for a standard type II SN.  The X-ray emission is concentrated in the central  $\sim 2~\pc$ of the 
remnant while no  emission has been seen  coming from the radio shell.  

These observations are compatible with the theoretical predictions for a typical SNR of that age
which is expected to be in an advanced cooling phase.   Since the non-thermal radio shell still 
exists, its present velocity must be  faster than the sound velocity in the ambient material.   
According to the authors of \cite{Chandra} this implies that the temperature  of the ionised gas 
 halo surrounding Sgr A East is substantially smaller than 70 K which is  consistent with the
IR observations.   The density of the gas halo is about $10^{3}~\cm^{-3}$ and it is 
almost homogenous barring the molecular cloud M-0.02-0.07 where it becomes as large as  $10^{5}~\cm^{-3}$ \cite{CoilHo}. This dense material is swept up by the expanding shell giving
 rise to a thick snowplough in the form of a dust ridge surrounding  the shell. The dust ridge is 
 substantially denser at the eastern edge, where Sgr A East is interacting with the cloud, giving rise 
 to  detectable molecular line emission  (molecular ridge) \cite{CoilHo}.
 
Soft X-ray and UV  and optical radiation are effectively  absorbed in such  a dense medium.
Even outside this opaque ridge, the flux of UV and X-ray photon density in the surrounding of the 
radio shell is much smaller than the IR radiation density.
 Indeed,  the density of UV photons required to explain the gas ionisation in the GC circumnuclear 
 disk   has been estimated to be $\simleq  10^{4}~\cm^{-3}$ \cite{Wolfire90}. 
 An even  smaller photon density has to be expected over most of Sgr A East region.  
 The X-ray luminosity of Sgr A East  observed by Chandra  implies  
a negligible photon density over most of such SNR.
The ionising source of the gas halo surrounding the GC was not identified by Chandra.  
According to the authors of \cite{Chandra} this source may be related with a past activity of
the super-massive black-hole in Sgr A* being presently in a quiescent phase.  
The  milliGauss  field strength measured by   Yusuf-Zadeh et al.  \cite{YZ96} in the 
 Eastern side of Sgr A East may  be  explained by the  adiabatic compression  of the pre-existing 
 field  with an  intensity $B_{\rm GC} \sim 10 - 100~\muG$ .
Polarimetric observations \cite{Novak2000},  which mapped the projected magnetic field direction in the Sgr A region, support this scenario. 
Therefore we think it is reasonable to assume a magnetic field strength of  few milliGauss in the molecular ridge crossing  M-0.02-0.07,  and an order of magnitude smaller field in the 
rest of Sgr A East non-thermal shell. 
The field coherence length should be  several parsecs in both cases. 

For what concerns particle acceleration, the weaker field present in the latter region is
 compensated by the  larger velocity of the western shell. 
Indeed, the theoretically predicted shock velocity in the gas halo is   
 $v_s \sim  6 \times 10^7~(t/5 \times 10^4 \yr)^{-0.6}~\cm \s^{-1}$ \cite{Lagage} 
while it is expected to be few  times smaller in the  $10^{5}~\cm^{-3}$  dense molecular cloud.
Barring the effect of  energy losses,  the maximal CR acceleration energy should be, 
therefore, roughly  the same over most of the Sgr A East radio shell.   
However, hadronic scattering in  the high  density cloud should  prevent nuclei acceleration up 
to the EeV by taking place in that region.
Acceleration up to the EeV  may be also prevented in the South-West  
arc,  less than 1 pc away from  the GC,  owing to pair production and nuclei PD energy losses
(see below) onto the $\sim 100$ K temperature  IR background present in that region. 

\section{Gamma rays from the GC}\label{gammas}

  Several experiments observed a $\gamma$-ray excess in the direction of  the GC. 
  
  EGRET satellite  \cite{EGRET} detected an excess  centred  on the GC  with an uncertainty of 
  $0.2^o$,  corresponding to a  projected lenght of  30 pc at the GC distance.  
 In the last few years several Cherenkov telescopes allowed to explore the GC emission at higher 
 energies.  WHIPPLE \cite{Veritas} and CANGAROO II \cite{Cangaroo} telescopes, 
also found a $\gamma$-ray  excess in direction of the GC  in the energy range $0.2 - 1~\TeV$.
 However, the angular resolution reached by these instruments was  not much better than 
 EGRET.  A breakthrough came only
 recently with the HESS telescope which, thank to its  stereo-graphic shower reconstruction
technique,  succeeded observing the GC emission up to 10 TeV with an angular resolution   
of 1' \cite{HESS}.  The background-subtracted differential flux measured by HESS is
$\displaystyle F(E) \simeq (2.76 \pm 0.33) 
  \times 10^{-15}   (E/1~\TeV)^{-2.2}~\cm^{-2} \s^{-1} \GeV^{-1}$ in the energy range 
  $165~\GeV -  6~\TeV$ \cite{HESS}.   The angular distribution of the signal is compatible
  with a source of size  $< 7$ pc  ($95 \% $ C.L.),  at the distance of the GC,  centred  on a point 
that, within errors, may coincide  either with Sgr A* or with the centre of Sgr A East remnant.
 The spectra observed by EGRET and HESS can hardly be connected so that the sources
 responsible for the two signals are probably different.  Since the excess observed by 
 EGRET is within the field of view of HESS but is not observed by that telescope, 
 the source responsible for the EGRET signal (3EG J1746-2851)  must be very faint above 100 
 GeV.  For these reasons  we do not consider  3EG J1746-2851  as a probable  site for the
 acceleration of CRs up to ultra high energies.
  
 Several sources have been proposed to be responsible for the  excess observed by HESS.  
  Among them there are the super-massive black hole in Sgr A*, in the 
 galactic dynamical centre, a  clumped halo of annihilating supersymmetric dark matter 
 in the GC and Sgr A East.  
In our opinion, Sgr A East is the most plausible source for the following reasons:
 \  a) the spectrum of the GC emission observed by HESS is very similar to that of several
other SNRs as measured by HESS itself and by  other Cherenkov telescopes;  
b)  no evidence of a high-energy cutoff was found up to 6 TeV  which disfavours
the dark matter related interpretation of the HESS signal;  c)  the $\gamma$-ray angular distribution
detected by HESS is slightly  Eastwards elongated (see Fig.1 in \cite{HESS});  
d) the  $\gamma$-ray flux is compatible with the theoretical estimate based on the
assumption that  such radiation  is  a secondary product of  hadronic cosmic rays
accelerated in a  SNR  with the same properties of Sgr A East.     

Due to the relevance of the last issue for the aims of this paper
we summarise here  the main passages of such computation. 

Similarly to what done in \cite{FatMel03, Crocker} for 3EG J1746-2851, 
we assume that the high energy $\gamma$-rays  detected by HESS from the direction of the GC
are generated by the hadronic scattering of  high energy  nuclei (mainly protons) 
accelerated by Sgr A East onto  the ambient gas nuclei (mainly hydrogen).
The relevant process is $pp \rightarrow pp~ +~ {\rm n}_\pi \pi^{\pm,0}$ ($ {\rm n}_\pi$ 
is the pion multiplicity)  with the  neutral pions decaying into photons.  At high energies  the 
differential  $pp$ cross section  can be approximated by a scaling form  \cite{Blasi99} and the 
photon  emissivity can be written \cite{Bere97,FatMel03,Crocker} 
\begin{equation}
Q_\gamma(E_\gamma) \simeq   n_p(E_\gamma)  
\sigma_{pp}^{\rm ine}(E_0)  c \; n_H 
Y_\gamma(\alpha)~,
\end{equation}
  where $ Y_\gamma(\alpha)$ is the so-called `spectrum-weighted moment', or `yield', of
secondary photons.  In our case $Y_\gamma(2.2) = 0.103$ \cite{Bere97,Crocker}.   
 Reasonably, the mean gas density in the  Sgr A East region will not differ too much from
  that of the ionised gas halo surrounding the GC which, according to Chandra observations,  is  
  $\sim 10^3~\cm^{-3}$ (while the gas density is larger  than this value in the molecular/dust 
  ridge, behind the shell it is smaller).
   We assume equipartition between the total energy of the relativistic particles accelerated by 
 Sgr A East and the thermal energy of the X-ray emitting plasma  which is
 amounting to $10^{49}$ ergs \cite{Chandra}.  Then, by assuming a power law  proton spectrum,  $\displaystyle N_p(E) = 
 N_0 ~\left(\frac{E}{E_0}\right)^{-\alpha}$,  and imposing 
 the normalisation condition  
 \begin{equation}
\label{normalisation}
E_{\rm RC} = ~\int_{E_0}^\infty dE~ E  N_p(E) =  
\frac{1}{\alpha -2} N_0 E_0^2  ~.
\end{equation}     
we find that, for $\alpha = 2.2$ and $E_0 = 1 \GeV$,  the expected $\gamma$-ray  flux reaching the Earth is
 \begin{equation}
F_\gamma(E_\gamma) \simeq 5 \times 10^{-15} \;  
 \left(\frac{E_\gamma}{1~\TeV} \right)^{-2.2}
 \left(\frac{n_H}{10^3~\cm^{-3}}\right) \left(\frac{E_{\rm CR}}{10^{49}~\erg}\right) 
 \GeV^{-1} \cm^{-2} \s^{-1} ~.
\end{equation}
 
 This is less than a factor two larger than the flux measured by HESS which, by accounting for 
 the large uncertainties in the quantities involved in this computation, amounts to an 
 excellent agreement between the model predictions and the observations. 
 
 We conclude this section by observing that TeV $\gamma$-ray attenuation 
 due to pair-production scattering onto the IR photon background in  
 the Sgr A East region is negligible.  The mean IR photon energy is, in fact, 
 $\displaystyle 3 kT \sim 10^{-2} \left(\frac{T}{40~K}\right) ~\eV$, implying a
 $\gamma$-ray threshold energy of  about 20 TeV.  The effect may be significant only in 
 that portion of Sgr A East shell which is within $1~\pc$  from the GC where $T \sim 100$ K. 
  Gamma-ray scattering onto the UV and X-ray backgrounds present in that region  was also showed 
 to give rise to a negligible attenuation \cite{Crocker}.

\section{Maximal energies of nuclei}\label{Emax}

\subsection{Acceleration time scale}

In the first order Fermi acceleration process,  the acceleration time  scale coincides with the mean
 time taken by a particle  to cycle through the shock \cite{Lagage}. This is roughly given by 
\begin{equation}
t_{\rm acc}(E) \simeq \frac{4}{c} \frac{D(E)}{u_S}
\end{equation} 
where $D(E)$ is the diffusion coefficient and $u_S $ is the shock velocity.
  When the  particle energy is very high it is reasonable to assume Bohm like diffusion 
 giving the minimum value of  the diffusion coefficient $D(E) = \frac{1}{3} c R_L(E)$. 
The Larmor radius of  a nucleus with charge $Z$ is
\begin{equation}
\label{ Larmor}
R_L(E)   = \frac{E}{Z~eB} \simeq  
    \frac{1}{Z} ~ 10 ~ \left( \frac{E}{10^{18}~\eV}\right)
\left(\frac{B}{100~\muG}\right)^{-1} \;  \pc~.
\end{equation}
An  important condition to be fulfilled  is that $R_L(E_{\rm max})$ does not exceed the size of 
the SNR.
   
Using the previous expressions we find
\begin{equation}
\label{tacc}
t_{\rm acc}(E) \sim \frac{1}{Z}   \left( \frac{E}{10^{18}~\eV}\right)
\left(\frac{B}{100~\muG}\right)^{-1}  ~ \left( \frac{u_S/c}{10^{-3}}\right)^{-1} \; 
10^{12} ~\s ~.
\end{equation} 

Nuclei acceleration up to UHEs might be prevented by one of the following processes:
a) hadronic collisions onto the gas particles;  b) electron-positron pair-production 
by the  interaction with low energy photons; c) photo-disintegration onto  background photons.   The relative relevance of these processes depends on the physical
conditions in the acceleration region. 

\subsection{Energy losses due to hadronic collisions}

The gas density in the GC region is considerable larger than the mean ISM density.
Chandra observations \cite{Chandra} allowed to infer that  Sgr A East shell is expanding
through a ionized gas halo with mean density $n_H\sim 10^3~\cm^{-3}$. The  hydrogen 
density  can be as high as $\sim 10^5~\cm^{-3}$ in the molecular cloud M-0.02-0.07 .
  The nucleus-proton scattering time scale is
\begin{equation}
\label{tpp}
t_{\rm hadr} \simeq  \frac{1}{A^{2/3}~c n_H  K \sigma_{pp}} \simeq 
A^{-2/3}~10^{12} ~\left( \frac{n_H}{10^3 ~\cm^3}\right)^{-1}~\s~,
\end{equation}   
 where the $pp$ cross section $\sigma_{pp}(E) $ and the inelasticity $K$ are given in the Appendix.
 The $A^{2/3}$ factor in  (\ref{tpp}) is due to the grown of the
 effective surface taking part to the $pA$ interaction respect to the $pp$ scattering. 
 We can safely neglect here the weak energy dependence of $\sigma_{pp}$ and $K$.  
 A more detailed treatment will be performed while determining the spectrum of the
  secondary particles produced by the $pp$ scattering (see  Sec.\ref{neu_pp}).

\subsection{Pair-production energy losses}

\begin{figure}[!ht]
  \begin{center}
  \includegraphics[scale = 0.4]{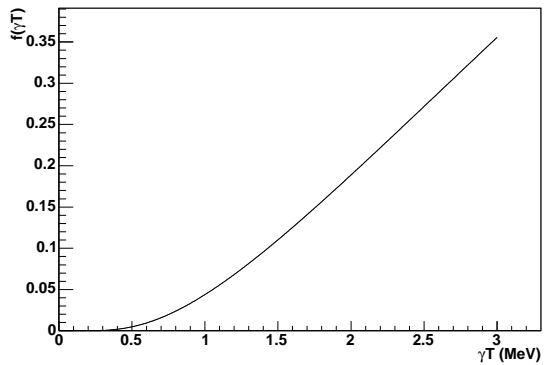}
   \end{center}
  \caption{The function $f(\gamma kT)$ is represented.}
  \label{fpair}
\end{figure}

As we discussed in Sec.\ref{observations}  the region where Sgr A East is
located is  pervaded,  besides by the hydrogen gas, by a background of black-body radiation at a 
temperature $\sim 40$ K.
Therefore, the  photon density in that region is considerably  larger than in the interstellar and intergalactic media where the  CMB  is the dominant radiation background.  
It is well known  in cosmic ray physics that a photon background give rise to significant energy
 losses of UHECRs.  
 Besides the PD, which in our case is a useful process and will be considered in  Sec.\ref{PD},
 the most relevant loss process for nuclei is  the pair production:   
 ($(A,Z) \gamma \rightarrow (A,Z) + e^+ e^-$) . 
This process has been studied in details in two seminal papers by Blumenthal \cite{Blumenthal:1970nn} and Chodorowsky et al. \cite{Chodo92}. 
 Applying the results found in  those papers we find 
\begin{equation}
\label{pair_rate}
\begin{split}
t_{\rm pairs}^{-1}  = & \frac{2 \alpha r_0^2 }{\pi^2 c\;^2
\hbar^3 }\frac{Z^2}{A}\frac{m_e}{m_p}  (kT)^3  f(\gamma T) \\
= & ~  \frac{Z^2}{A}  f(\gamma T)  
\left(\frac{T}{40~{\rm K}}\right)^3 \; 1.4 \times 10^{-15} ~ \s^{-1} 
\end{split}
\end{equation}
where $\gamma =  E/A m_N$.  The function
\begin{equation}
f(\gamma T) = \left(\frac{m_e c^2}{\gamma kT}\right)^3 ~
\int_2^\infty d\xi\;  \phi(\xi) \frac{1}{ e^{ {\xi m_e c^2}\over {2 \gamma kT}} - 1 }~,
\end{equation}
 where  $\phi(\xi)$ is defined in \cite{Chodo92},  is plotted in Fig. \ref{fpair}.

Therefore, it follows from Eq.s (\ref{pair_rate}) and  (\ref{tpp}) that in Sgr A East
pair-production losses of ultra-relativistic nuclei are subdominant respect to hadronic losses.
Photo-pion production also does not play any  significant role in  the energy  range of interest for
this work.

\subsection{Maximal energy}

An order of magnitude estimate of the maximal energy reachable by diffusive shock  acceleration
can be determined by imposing the condition 
\begin{equation}
\label{t_cond}
t_{\rm acc}(E) = {\rm min}\left\{  t_{\rm loss}(E),~t_{\rm SNR} \right\}~,
\end{equation}
where $t_{\rm acc}(E) $, $ t_{\rm loss}(E)$ and $t_{\rm SNR}$ are respectively
the characteristic acceleration time, energy loss time and the age of the SNR. 
As we showed in the above, energy losses in the Sgr A East region are dominated by 
hadronic scattering with a characteristic time given by Eq. (\ref{tpp}).
Sgr A East  age is 
\begin{equation} 
\label{Sgr_age}
t_{\rm SgrAEast} \sim 5 \times 10^{4} ~\yr \simeq 10^{12}~\s~.
\end{equation}
which, especially for heavy nuclei, is larger than  $t_{\rm hadr}$.

Therefore by substituting (\ref{tacc}) and (\ref{tpp}) into (\ref{t_cond}) we find 
that the maximal energy reachable by shock acceleration in Sgr A East is roughly given by 
\begin{equation}
\label{ E_max}
E_{\rm max} \simeq    ~  \frac{Z}{A^{2/3}} ~ \left( \frac{u_S/c}{10^{-3}}\right)  
\left(\frac{B}{100~\muG}\right)   \;~ 10^{18}~\eV~.
\end{equation} 
We conclude that  the acceleration of nuclei  up to the ankle of the CR spectrum  in 
Sgr A East is  plausible.
According to Croker {\it et al.},  even larger energies may be reached in that object if
the magnetic field has a component non-perpendicular to the shock.

\section{Neutron production by nuclei photo-disintegration  onto the IR background}\label{PD}

To determine the production rate of neutrons by the photo-disintegration (PD) of high energy 
nuclei interacting with a heat bath of  photons we apply the results of Puget, Stecker and 
Bredekamp  \cite{Puget:76}.
Following what done in \cite{Puget:76}, the photo-disintegration rate for a nucleus of 
atomic number $A$ (we consider only stable nuclear species) releasing $i$ nucleons  
can be written as
\begin{equation}
\label{rate_gen}
R_{A, i} = \frac{1}{2 \gamma^2} \int_0^\infty d\epsilon \epsilon^{-2} n(\epsilon) 
\int _0^{2\gamma \epsilon} d\epsilon' \epsilon' \sigma_{A, i}(\epsilon' )
\end{equation}
where $\gamma$ is the Lorentz  factor of the nucleus, and $\epsilon$, $\epsilon'$ the photon 
energy in the observer and nucleus rest frames respectively.   
 In general, the  PD cross section is well approximated by the sum of two terms:  a term
describing one- and two-nucleon PD, which is dominated by the giant dipole resonance,
and a non-resonant term which is relevant for the multi-nucleon emission only. 
 For $\gamma kT  \ll 30 ~\MeV$, which is the case in the present context,  the latter term gives a 
 negligible contribution. The function
\begin{equation}
\sigma_{A, i}(\epsilon' ) \simeq \sum_{i = 1}^2
\theta(\epsilon' - \epsilon^{\rm th}_{A,i}) \theta(30~\MeV - \epsilon' )
 ~ W_{A,i}^{-1} \xi_{A, i} \Sigma_d   \exp\left[ - \left(\frac{\epsilon' - 
 \epsilon'_{o ~A,i}}{\Delta_{A,i}}\right)^2
 \right]
 \end{equation}
 was found to provide a good fit of the experimental data \cite{Puget:76}.
  The definitions of the parameters $\epsilon^{\rm th}_{A,i}$, $\epsilon'_{0, i}$, $\Delta_{A,i}$
  and  $\xi_{A, i}$ are given in  \cite{Puget:76,Stecker:99}.  In Tab.\ref{tab1} we report their 
  values for the PD  of most abundant nuclides  with the emission of only one nucleon.  For the 
  threshold energies   $\epsilon^{\rm th}_{A,i}$ we use the values given in Stecker and Salamon 
 \cite{Stecker:99}  which we also report in the same table.
The strength of the PD reaction, i.e., the integral of cross section over the relevant energy
interval, is  $\displaystyle \Sigma_d = {{(A -Z)Z}\over{A}} 5.98 \times 10^{-26} ~\MeV \cm^2$.
 The normalisation constant is given by 
 \begin{equation}
W_{A,i} =  \sqrt{\frac{\pi}{8}} \left[ {\rm erf}
 \left(\frac{30~\MeV -  \epsilon'_{0,~A, i}}{\Delta_{A,i}/\sqrt{2}}\right) +
 {\rm erf}\left(\frac{\epsilon'_{0~A,i} - \epsilon^{\rm th}_{A,i}}{\Delta_{A,i}/\sqrt{2}}\right) 
 \right]~.
\end{equation}
 
  \begin{table}[!hb]
\begin{center}
\begin{tabular}{|c|c|c|c|c|c|c|}
\hline Nuclear species &  A & Z & $\epsilon^{\rm th}_{1}$ &$\epsilon'_{0~1}$
   &$\Delta_{1}$&$\xi_{1}$\\
\hline Fe & 56 & 26 & 11.2  & 18 & 8 & 0.98\\
\hline O  & 16 & 8 & 15.7  & 24 & 9 & 0.83\\
\hline C  & 12 & 6 & 18.7 & 23 &  6 & 0.76 \\
\hline He  & 4 & 2 & 20.6  & 27 & 12 & 0.47\\
\hline
\end{tabular}
\end{center}
\caption{PD cross section parameters for the most abundant nuclides  as given in \cite{Puget:76,Stecker:99}. The $\epsilon$'s  and $\Delta$ are in MeV.}
 \label{tab1} 
 \end{table}
 
 We find  convenient to express the photo-disintegration rate in the form:  
 \begin{equation}
\label{PDrate}
R_{A, i} (\gamma T)   \simeq R_0 ~ \xi_{A, i}  \left (\frac{T}{40~{\rm K}} \right)^3
   \Phi_{A, i}(\gamma T)
\end{equation}
where
\begin{equation}
R_0  
= \frac{1}{2 \pi^2 \hbar^3 c^2}~ (5.98\times 10^{- 26}~\cm^2)~
\left( k~ 40~{\rm K}\right)^3 
\simeq 5 \times 10^{-10} ~\s^{-1}~.
\end{equation}
 The function $\Phi_{A,i}$ is defined by
 \begin{equation}
\label{Phi}
\Phi_{A,i}(x)  \equiv  (W \Delta)_{A,i}^{-1} ~\frac{(A-Z) Z}{A} \;  x^{-3} 
\int_{\epsilon^{\rm th}_{A,i}}^{15~\MeV} d{\tilde \epsilon} \left( e^{-{\tilde \epsilon}/{x}} - 1 \right)^{-1} J_{A,i}({\tilde \epsilon})~,
\end{equation}
where 
\begin{equation}
J_{A,i}({\tilde \epsilon}) \equiv \int_{\epsilon^{\rm th}_{A,i}}^{2 \gamma {\tilde \epsilon}}
\displaystyle
dy~ y \exp\left\{ -2\left( \frac{y - \epsilon'_{0~A, i}}{\Delta_{A,i}} \right)^2\right\}~.
\end{equation}
 We verified that further contributions to the PD rate which were considered in \cite{Puget:76} 
 are subdominat in our case.   

\begin{figure}[!ht]
  \begin{center}
  \includegraphics[scale = 0.3]{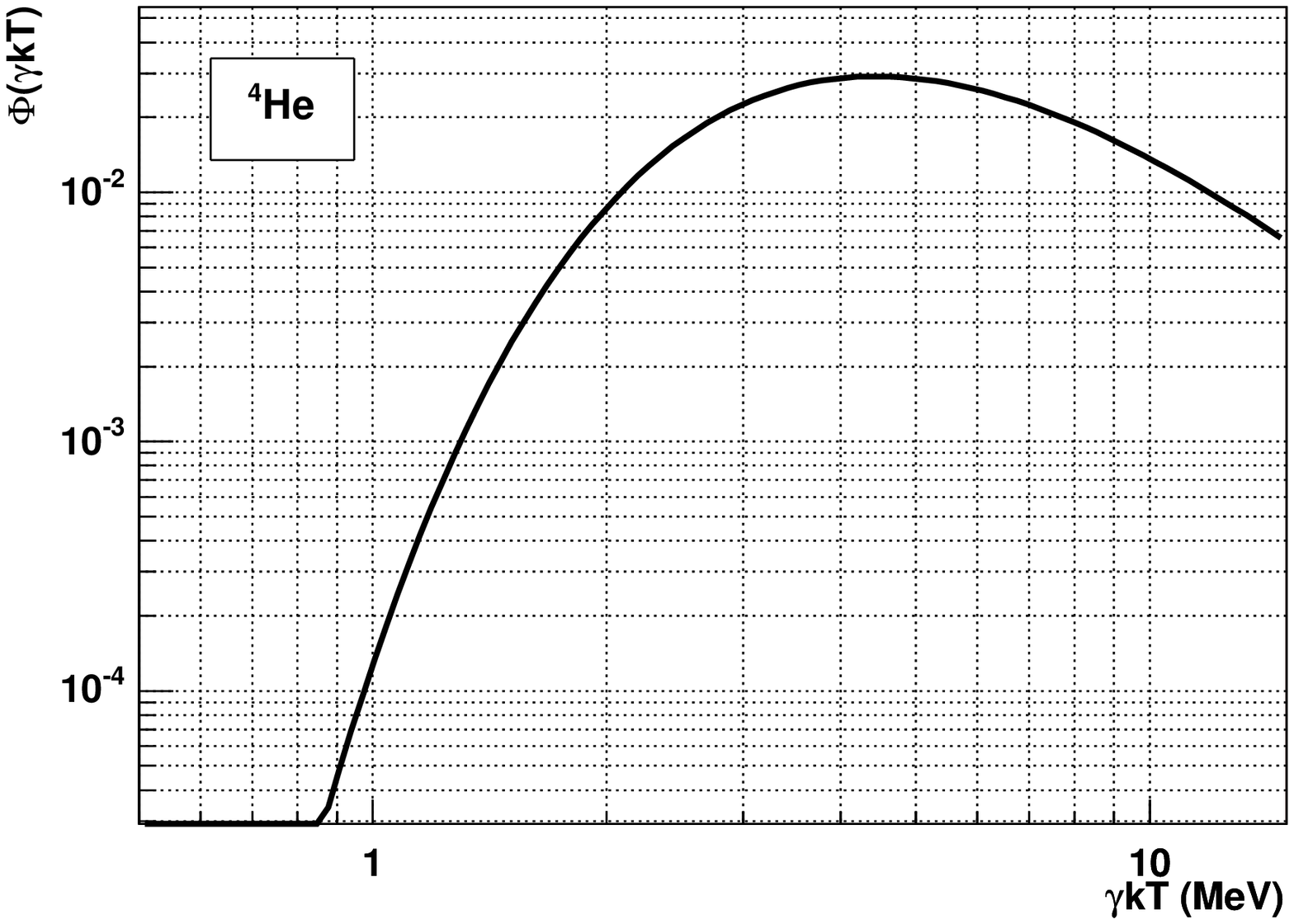}
  \includegraphics[scale = 0.3]{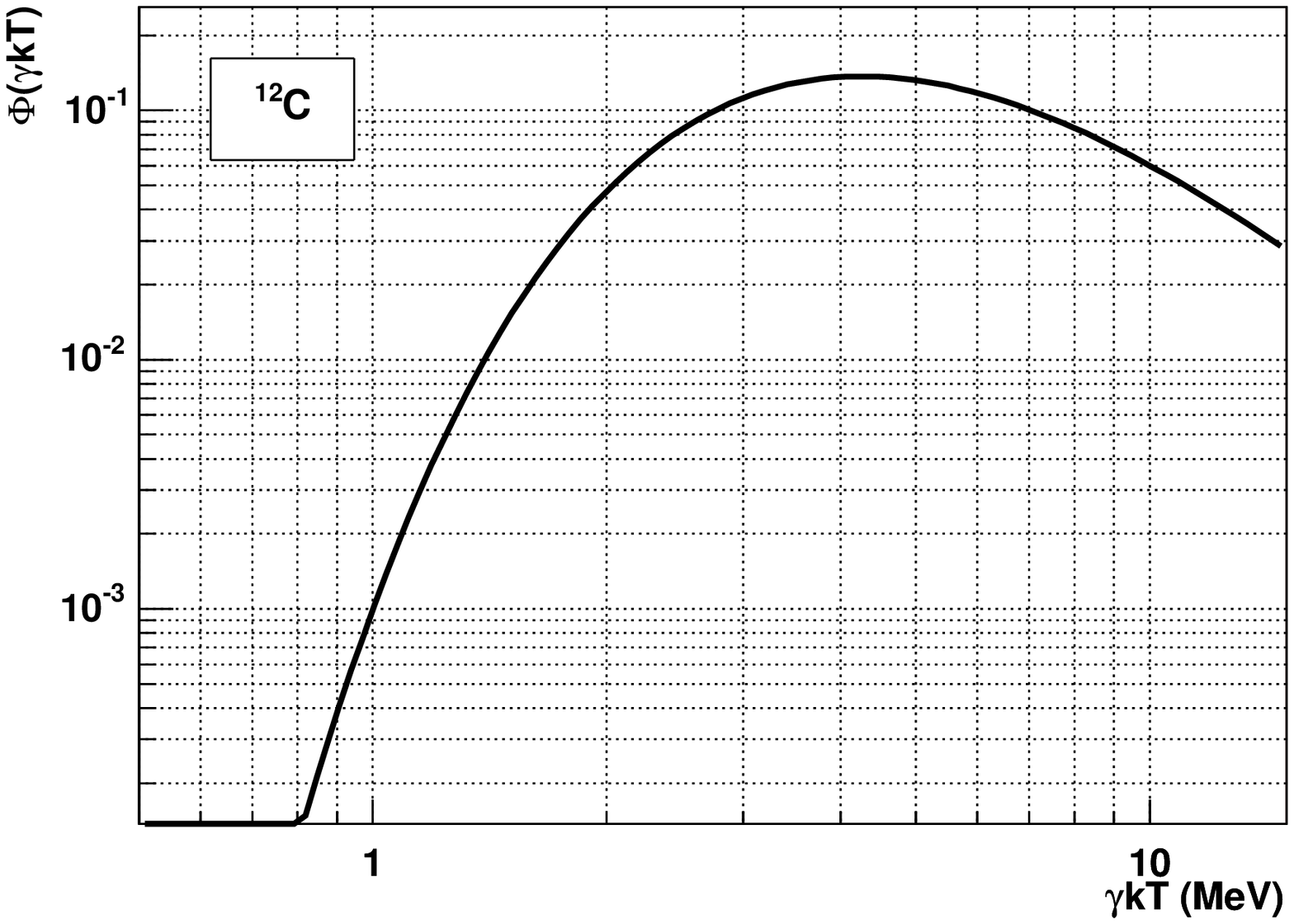}
  \includegraphics[scale = 0.3]{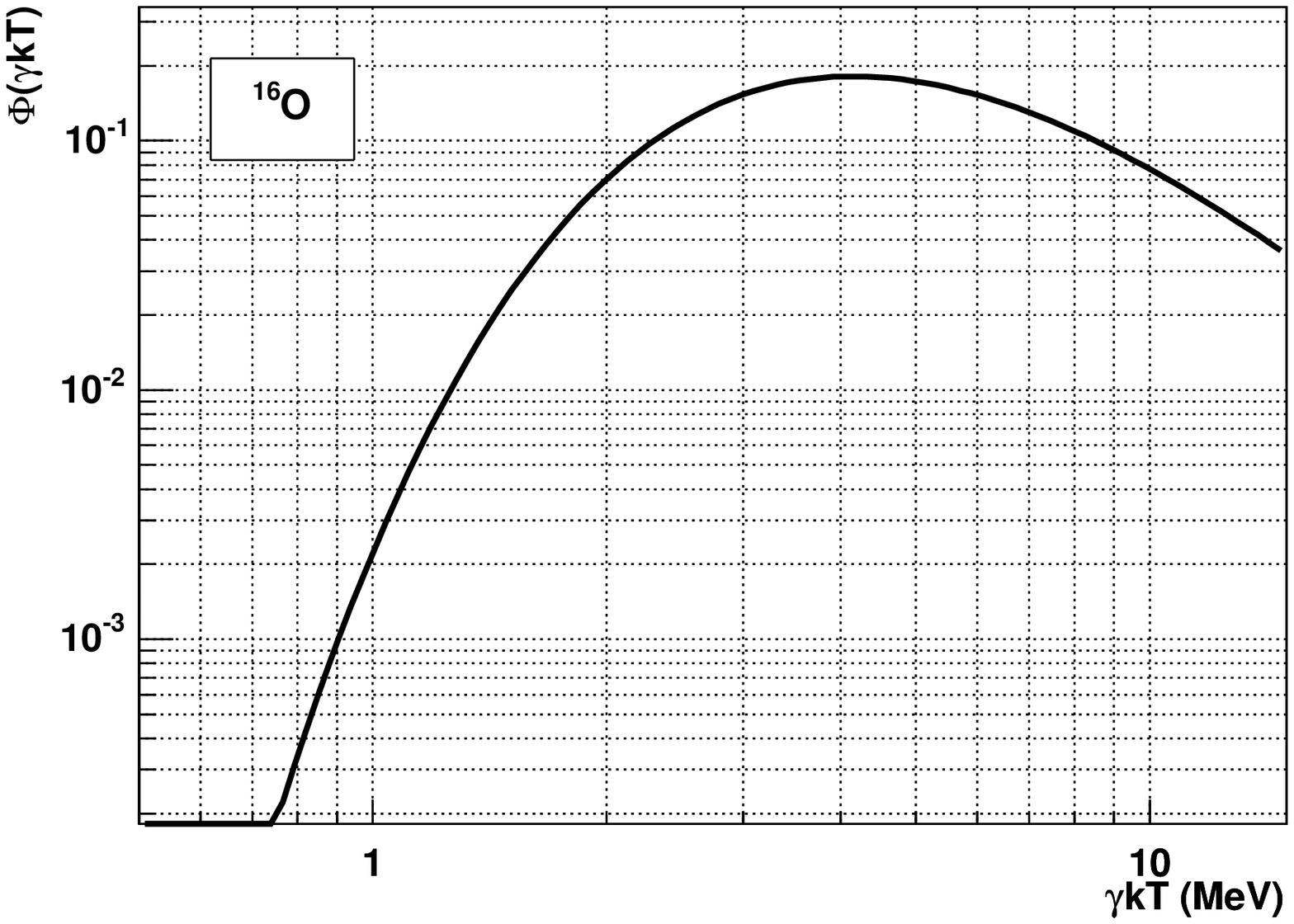}
  \includegraphics[scale = 0.3]{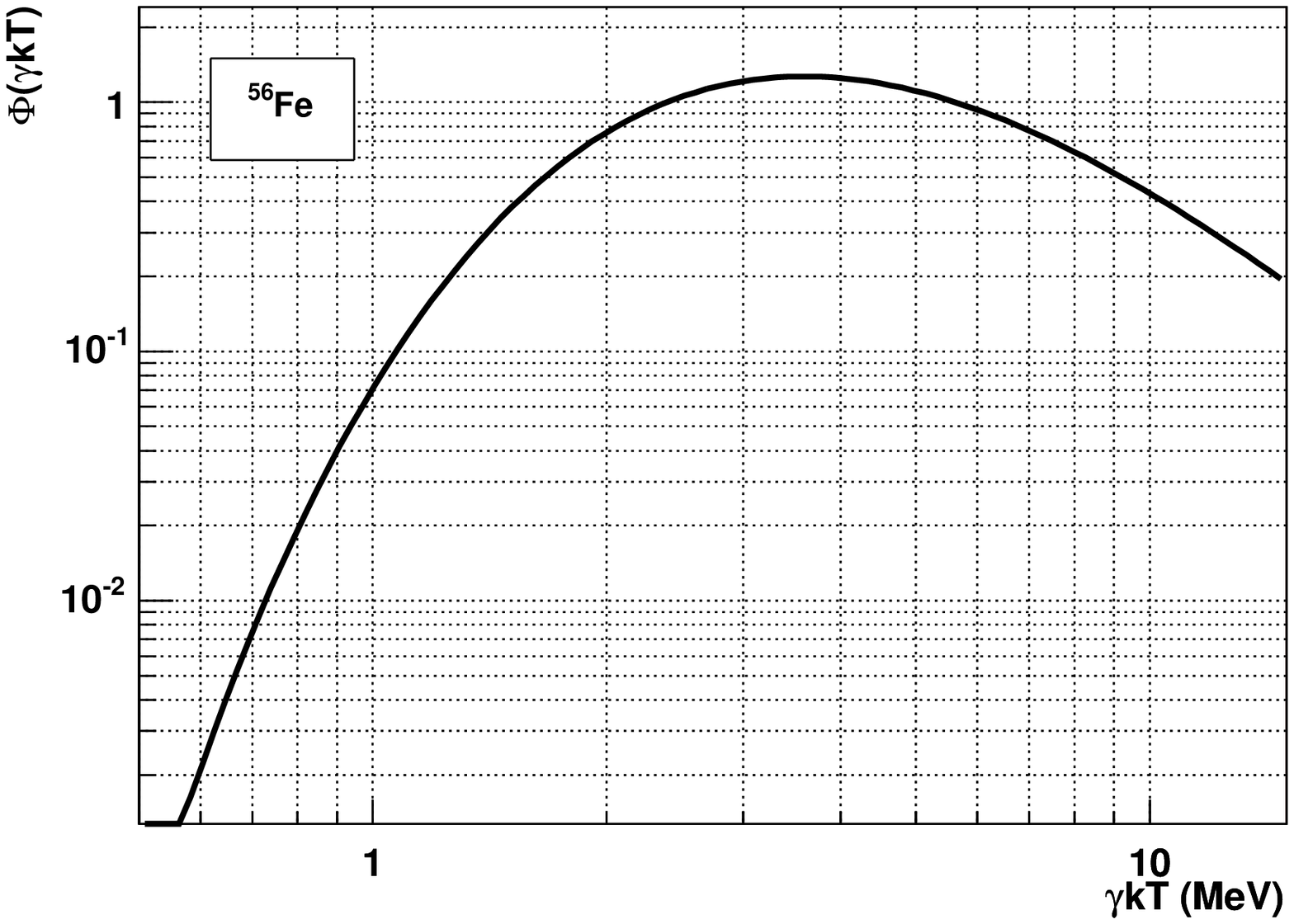}
   \end{center}
  \caption{The function $\Phi_{A,1}(\gamma kT)$ is represented for  $^4$He, $^{12}$C,  $^{16}$O  and $^{56}$Fe.}
   \label{graphPhi}
\end{figure}

In Fig. \ref{graphPhi} we represent the function $\Phi_{A,1}$ for several nuclear species.
Interestingly,  we found that  for all  relevant nuclei the  photo-disintegration rate is peaked at  
$ \gamma k T \approx   4 ~\MeV$. We will see that this property  has relevant consequences
for the spectrum of the secondary neutrons.   
In terms of the nucleus energy the peak condition reads
\begin{equation}
\label{E_A_max}
E_A^{\rm peak} \simeq A \times 10^{18}~\left( \frac{T}{40 ~{\rm K}} \right) ^{-1}~\eV~. 
\end{equation}
The maximal value of $\Phi_{A,i}$ is different for different  nuclear species.  We consider here only those species which are expected
to have the largest  abundance in   Sgr A East. The reader can see from Fig.\ref{graphPhi} that, 
among the most abundant elements, iron  ($^{56}$Fe)  nuclei have  the highest maximal value of 
the PD rate.  
However, due to grow of energy losses with $A$ (see Eq.{\ref{tpp}),  we do not expect heavy
 nuclei to reach $E_A^{\rm peak}$. Furthermore,  even disregarding hadronic energy losses,
 a too rapid  PD (note that the maximal PD rate of $^{56}$Fe is  much larger than  the inverse age
  of Sgr A East)  would itself prevent Iron  nuclei reaching  $E_A^{\rm peak}$.  
 This is not the case for  $^4$He nuclei 
 having a PD peak rate  which is almost coincident with the  inverse of the Sgr A East age.  
 We assume here that the physical conditions in Sgr A East  (namely the local magnetic field 
 intensity and geometry nearby the shock) allow the acceleration 
 of $^4$He nuclei up to few  EeV, corresponding to a value of $\Phi_4$ of 
 $(2 \div 3) \times 10^{-2}$. As we showed in Sec. \ref{Emax} such conditions are not  unrealistic.
 
 In order to evaluate the flux of secondary neutrons reaching the Earth we need to estimate
  the relative abundance of the main nuclear species in the  SNR shell.   
Since we are assuming that all nuclei  are accelerated by the same mechanism under the same 
physical conditions, we expect them to share the same power-law spectrum, namely 
$N_A(E) =  f_A  N_{0} \left( \frac{E}{E_0}\right)^{-\alpha}$, where $f_A$ parametrizes the 
relative abundance of nuclear species in the SNR shell.  Although, as we discussed in 
Sec.\ref{Emax},  the spectra of different nuclear  species  may  have different UV cut-off energies, 
for $\alpha > 2$ this has no  significant effect on the CR energetic. $N_0$ can be 
readily determined by imposing the normalisation condition (\ref{normalisation}).

Assuming a uniform cosmic ray density in the acceleration region, the  neutron emissivity  is 
therefore
\begin{equation}
\label{Qn_PD}
Q(E_n)  = \frac{1}{2~V}  \sum_A \sum_{i = 1}^2   i 
f_A   (\alpha -2)   E_{\rm CR} 
\left(\frac{A E_n}{E_0}\right)^{-\alpha} ~  
{\rm min}\left\{ R_{A, i}, ~ t_{\rm SNR}^{-1} \right\}
\end{equation}
where $m_N = (m_p + m_n)/2$ and $V$ is the volume of the emitting region. 
We assumed a probability $1/2$ for  the expelled nucleon to be a  neutron. 
 We used here that any neutron expelled during the photo-disintegration  carries, in average,
 a fraction $1/A$  of the parent nucleus  energy. 
 The two nucleon emission process is subdominant and we neglegt it in the following.
 The minimum condition in (\ref {Qn_PD})  is required since for 
 $ R_{A}(\gamma T) > t_{\rm SNR}^{-1}$  
 any further acceleration  is effectively inhibited by  the PD energy losses. 
 
According to Chandra observations,  Sgr A East metallicity is about four time that 
of the Sun. As a consequence we estimate the iron/hydrogen  relative abundance to be
$f_{56} \simeq 10^{-3}$.  The PD condition ($R(56) \sim t_{\rm SNR}^{-1}$) is reached
for $\gamma kT \sim  0.7 ~\MeV$ corresponding to a neutron energy $\sim 2 \times 10^{17}~
\eV$. 
Carbon and Oxigen should have a comparable abundance and a smaller peak rate
so that their PD gives rise to neutrons of energy  $\sim {\rm few} \times 10^{17}~\eV$. 
These neutrons however are hardly detectable due to their smaller time-life and the  
small abundance of the corresponding primary nuclei.
The most significant contribution to the neutron flux reaching the Earth 
should come from the PD of $^4$He nuclei. 
Indeed $f_4 \simeq 0.1$ in Sun, and it is presumably even larger in the remnant of a type II SN
(unfortunately  $^4$He is quite difficult to observe).
Furthermore, as we showed in the above, for the $^4$He the PD 
condition is  reached just at the energy where the PD rate takes its maximal value
so that, as follows from  Eq. (\ref{E_A_max}),  the flux of secondary neutrons will  peak at the energy 
\begin{equation}
\label{n_energy}
E_n^{\rm peak} \simeq  \frac{E_A^{\rm peak}} {A}  =  1 \times 10^{18} ~
\left( \frac{T}{40 ~{\rm K}} \right) ^{-1}~\eV~. 
\end{equation}
This is just the energy required for the neutron flux to be not suppressed significantly
by the decay on the way from the GC to the Earth. 

Using (\ref{Qn_PD}) and  accounting for the finite lifetime of neutrons we finally  find that for
 $\alpha = 2.2$ the expected neutron flux reaching the Earth from  Sgr A East is
\begin{equation}
\begin{split}
\frac{dF(E_n)}{dE_n}
&\simeq  2 \times 10^{-27} ~\left(\frac{f_A}{0.1}\right)  
\left( \frac{R_0 \xi_A \Phi_{A}(E_n, T)}{10^{-12}~\s^{-1}}\right)
 \left( \frac {E_{\rm CR}}{10^{49}~\erg}\right)
\left(\frac{d}{8~\kpc}\right)^{-2}  \\
& \left (\frac{T}{40~{\rm K}} \right)^3
\exp\left(-{d ~m_n\over E_n ~c  \tau_n}\right)
\left(\frac{A E_n}{10^9~\GeV}\right)^{-2.2}
~\GeV^{-1} \cm^{-2} \s^{-1}~.
\end{split}
\label{flux_PD}
\end{equation}
Note that, since the neutron mean time-life is $\tau_n = 887~\s$ \cite{PDG},   
the neutron survival probability for $E_n = 10^{18}~\eV$ and $d = 8~\kpc$ is
$P_{\rm sur} = \displaystyle  \exp\left(-{d ~m_n\over E_n ~c  \tau_n}\right) \simeq 0.43$.

The spectrum of neutrons produced by the PD of $^4$He nuclei is represented in Fig.\ref{neu_PD}.  While below  1 EeV the flux is exponential suppressed due to the
combined effect of the PD threshold and the neutron decay, above that energy 
the neutron spectrum is a power-law of index $\sim 5$ as an effect of the convolution 
of the primary nuclei spectrum with the $E^{-3}$    high energy behaviour of the PD rate. 
The value $f_4 = 0.1$ was assumed drawing   Fig.\ref{neu_PD}.  
This is a quite conservative assumption, as  the relative $^4$He abundance in Sgr A East is likely 
to be few times larger.
\begin{figure}[!ht]
  \begin{center}
  \includegraphics[scale = 0.5]{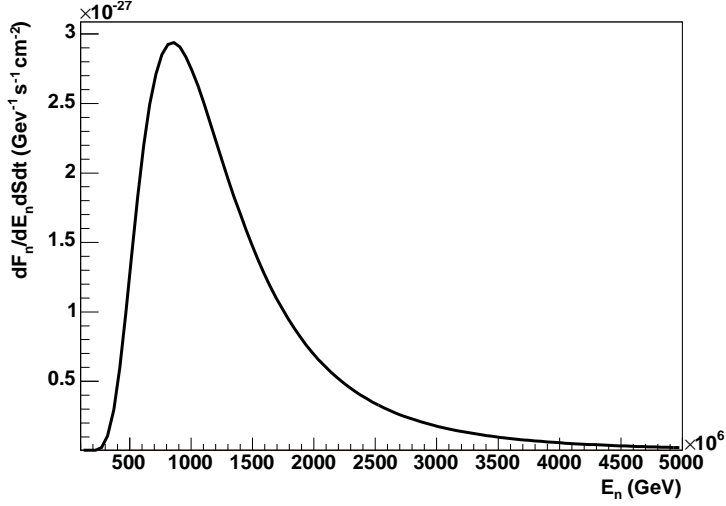}
   \end{center}
  \caption{The differential flux reaching the Earth of neutrons produced by 
   the photo-disintegration of $^4$He nuclei in Sgr A East}
   \label{neu_PD}
\end{figure}

We conclude this section by observing that PD of $^4$He nuclei  onto the UV and X-ray backgrounds in the surrounding of Sgr A East cannot prevent  their acceleration up to UHEs.
In fact,  in Sec.\ref{observations} we argued that the number density of photons in that 
energy range is $n_{\rm UV} \simeq n_{X} \simleq 10^{4}~\cm^{-3}$. This is smaller
than the photon density of the IR background at 40 K  by more than two orders of magnitude.
We showed in the above, that the PD rate does not depend on the absolute value of the nucleus 
energy but only on the product of its Lorentz factor with the photon energy. 
As a consequence,  the maximal PD rate due to UV and X-ray is the same of (\ref{PDrate}) but 
for the re-scaling  factor $n_{\rm UV, ~X}/n_{\rm IR} \simeq 10^{-2}$.  
 In the case of   $^4$He nuclei this quantity is smaller  than Sgr A East age by the same amount.
  
\section{Neutrons  from pp  collisions}\label{neu_pp}

Besides the PD process discussed in the previous section, EeV neutrons 
might be also be produced  by the collision of  UHE  nuclei with the  dense  gas in the GC 
environment.  The relevant neutron production channel is the charge exchange $pp$ inelastic collision $pp \rightarrow p n + {\rm n}_\pi \pi$,  where $ {\rm n}_\pi$ is the pion multiplicity. 
 
The neutron emissivity is
\begin{equation}
\label{Qn_pp_gen}
Q_n^{pp}(E_n) = c \; n_H \int_{E_{th}(E_n)} dE_p\,
n_p(E_p) {d\sigma^{\rm inel}(E_n, E_p)\over dE_n}
\end{equation}
where $\displaystyle  {d\sigma_{pp}^{\rm inel} (E_n, E_p)\over dE_n}$ is the $pp$ differential  
inelastic cross section. This quantity  is experimentally undetermined for $E_p$ larger than few 
hundred GeVs.
We handle this problem by combining Monte Carlo  simulations  with an analytical calculation  
 based on the scaling properties of the differential cross section at  high energy.
 In the Appendix  we find
 \begin{equation}
 \label{Qn_pp}
Q_n^{pp}(E_n) = c \; n_H  n_p(E_n) \sigma_0  0.24~ \frac{4}{5}  
 \left[ I_1(\alpha) + I_2(\alpha)  \ln\left(\frac{E_n}{E_0}\right)  \right]
\end{equation}
where $\sigma_0 \equiv \sigma_{pp}(E_0)$ and the functions $I_1(\alpha)$ and  $I_2(\alpha)$ are given in (\ref{Ialpha}). 
 
 Respect to a similar  approach  followed  by the authors  in \cite{Crocker}, our 
 scaling function  provides a better fit of the MC data at high  values of $E_n/E_p$ and 
 takes properly into account  the grown of the  cross section at very high energies. 
 As we did in the previous sections  we assume here that the energy spectrum of protons 
 accelerated by the SNR is a power law of index $\alpha$. 
 
Then,  by assuming $\alpha = 2.2$ and 
accounting  for the finite lifetime of the neutron,   we find that the differential neutron flux 
due to $pp$ scattering  which should reach  the Earth  is:
\begin{eqnarray}
\nonumber
F_n(E_n) &\simeq&  8 \times 10^{-28}   
 \left(\frac{E_n}{1~\EeV} \right)^{-2.2}
 \left(\frac{n_H}{10^3~\cm^{-3}}\right) \left(\frac{E_{\rm CR}}{10^{49}~\erg}\right) \\
 && \left(\frac{d}{8~\kpc}\right)^{-2}   \exp\left(-{d ~m_n\over E_n ~c  \tau_n}\right)
\  \GeV^{-1} \cm^{-2} \s^{-1} ~.
\label{flux_pp}
\end{eqnarray}
This is represented in Fig.\ref{n_pp}.

\begin{figure}[!ht]
  \begin{center}
  \includegraphics[scale = 0.5]{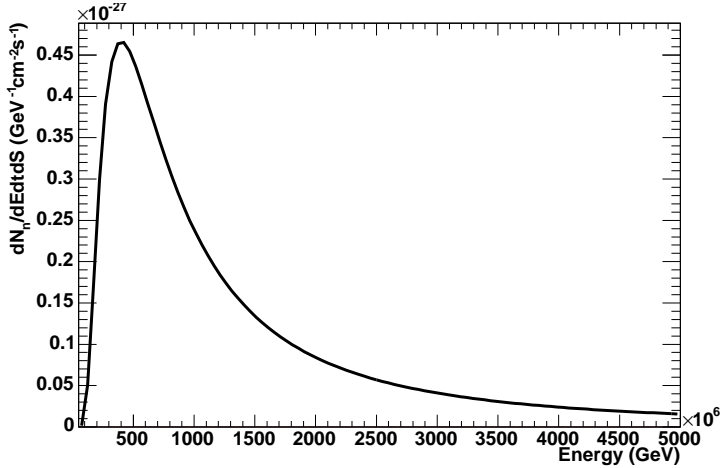}
   \end{center}
  \caption{The differential flux reaching the Earth of neutrons produced by 
  $pp$ collisions. }
  \label{n_pp}
\end{figure}

A comment is in order here.  In the appendix we showed  that the mean value of the 
neutron elasticity ($\langle x \rangle \equiv \langle E_n/E_p \rangle$) is $\sim 0.05$. 
Therefore, the maximal proton energy  must be larger than $2 \times 10^{19}~\eV$  to give
rise to unsuppressed neutron flux. 
In the derivation in the above we didn't account for a UV cut-off in the proton 
energy spectrum implicitly assuming that this  is far above  the EeV.  This assumption, however, 
is probably unrealistic even for Sgr A East (see Sec.\ref{Emax}). 
As a consequence, a suppression of the neutron flux due to $pp$ scattering  
 has to be expected respect to the estimate given in (\ref {flux_pp}) unless Sgr A East 
 is an extremely energetic proton accelerator.

\section{Expected signal at AUGER}\label{AUGER}
 
The Pierre AUGER Observatory  is an Extensive Air Shower detector under construction 
in Argentina \cite{Auger}. When completed, it will consist of 1600 Water Cherenkov 
elementary Detectors (WCDs) covering an area of about $3000~\km^2$ overlooked by
a set of four florescence telescopes.  An Engineering Array of WCDs has been in operation since 2002  and the detection of several showers with energy of the primary  $\sim 10^{18}~\eV$ has already announced at the ICRC 2003 \cite{Ghia}.  
The AUGER acceptance will be $A \Omega \approx 4700~\km^2
\sr$, for a zenith angle $\theta < 45^o$.   
The array trigger efficiency at the EeV has been estimated to be $\eta \simeq 0.25$ \cite{Ghia}.

The geographic position of AUGER is ideal for observing a UHECR excess in the direction of the GC.
Since the angular size of Sgr A East is few primes, the neutron emission from this object should 
give rise to a  point-like signal  in AUGER.
\footnote{
In principle, EeV protons produced nearby the Earth by  neutron decay  may give rise to a smeared signal superimposed to the point-like excess \cite{Medina,Roulet03}.
This secondary signal, however,  is probably undetectable.}
More precisely,  the expected  signal  should be seen  under the form of a CR excess  in the angular 
bin corresponding to the position of the source.  The angular bin covers a  solid angle 
$\omega = \pi \delta^2$, where $\delta$ is the angular resolution of the detector. 
The AUGER angular resolution $\delta$ is expected to be around $1^o$ at $10^{19}-10^{20}~\eV$ (implying $\omega \simeq 1 \times 10^{-3}~\sr$) but  it $\delta$  will
probably be $2^o-3^o$  (in the final configuration) around the EeV \cite{Auger}. 

The isotropic background of ordinary CRs gives rise to an unavoidable noise.
The CR spectrum in the energy range $4 \times 10^{17} < E < 6.3 \times 10^{18}~\eV$ is
\cite{Nagano}
\begin{equation}
\label{CR_flux}
J_{\rm CR}(E) = ( 9.23 \pm 0.65) \times 10^{-28} 
\left(\frac{E}{6.3 \times 10^{18}} \right)^{- 3.20\pm 0.05}
(\GeV ~\sr~ \cm^{2}~ \s)^{-1} ~.
\end{equation}
  Therefore the maximal (at $95 \% $C.L.) differential flux around $10^{18}~\eV$  
incident onto the expected AUGER angular bin  is 
\begin{equation}
F_{\rm CR}^{\rm max}(E) \simeq 3 \times 10^{-27} 
\left(\frac{E}{10^{18}~\eV} \right)^{- 3.3}~
\left( \frac{\delta}{2.5^o}\right)^2~
 \GeV^{-1} \cm^{-2} \s^{-1} ~.
\end{equation}

It is comforting  that the  expected neutron flux from Sgr A East produced  by the
 photo-disintegration of $^4$He nuclei (see Eq. \ref{flux_PD})  is comparable to the isotropic 
 CR flux around the EeV incident over the AUGER angular bin. 
As a consequence,  AUGER may observe a significant UHECR excess  in the 
angular bin corresponding to the position of the GC.
 
 As a benchmark,  we give here  the expected rates of  showers  incident over an angular bin 
  of $2.5^0$,  in the energy range 
$1 \times 10^{18} < E < 3 \times 10^{18}~\eV$,  due
to the PD and $pp$ neutrons respectively (accounting for the AUGER trigger efficiency) 
\begin{equation}
{\dot N}_{\rm PD} \simeq  132~\yr^{-1}\quad  {\dot N}_{pp} \simeq  72~\yr^{-1}~,
\end{equation}
 ${\dot N}_{\rm PD} $ has been computed assuming the $^4$He relative abundance 
 $f_4 = 0.2$ (see the discussion at the end of  Sec.\ref{PD}). 
 These rates have to be compared  with a CR background rate in the same angular and energy bins
which is  
\begin{equation}
{\dot N}_{\rm CR} \simeq  440~\yr^{-1}~.
\end{equation}
Since the expected signal is a sizeable fraction of the background  a significant excess 
should be detectable in the direction of the GC. 

We conclude that, under the conditions that we assumed in this paper,  the signal due to 
  EeV  neutrons produced by  Sgr A East should not be missed by AUGER.  
 
\section{Discussion}\label{discussion}

The possibility, discussed in this paper, of testing if nuclei acceleration takes place 
in  Sgr A East  up to  Ultra  High Energies  ($E > 10^{18}~\eV$) by looking for EeV neutrons
coming from the GC direction, was also   considered in a recent paper by  Crocker {\it et al.} \cite{Crocker}. 
Our work, however,  differs from that of Crocker  {\it et al.} in a number of relevant aspects. 

First of all, the aim of Crocker  {\it et al.} was to interpret the anisotropy  in the CR 
angular distribution  observed, around the EeV,  by the AGASA \cite{Hayashida:1998qb} and the 
SUGAR \cite{Bellido:2000tr} experiments in a direction relatively close to that of the GC.
This is not our attitude. Indeed, even accepting AGASA and SUGAR results as evidences of real 
CR anisotropies, which is tricky due to their  rather low  statistical significance,  it is, in our opinion,
  rather hazarding to identify Sgr A East as the source responsible for these signals.   
 Indeed, neither the AGASA nor the SUGAR UHECR excesses are in the direction of the very GC.
The GC is not even in the ``field of view" of AGASA being below its geographic horizon. 
\footnote{We have to note here that  an extended signal due to the  protons produced by neutron  decay has to be expected \cite{Medina,Roulet03}.  Although, this signal might explain
the anisotropy observed by AGASA in terms of a neutron source located in the GC, the 
required neutron luminosity of such source, which was determined in \cite{Roulet03}, is too high to be compatible with its possible  identification with Sgr A East.} 
 While SUGAR is in the right geographic position for observing the GC, no excess was observed by 
 this experiment from the true centre of the Galaxy. The SUGAR peak  was $7.5^o$ away from 
 the GC  and it does not even appear to be centred on the galactic plane.  This offset is too
 large to be due  to a possible systematic pointing error of the experiment.
 For these reasons   we think that Sgr A East cannot be responsible  for the AGASA and SUGAR anisotropies.  It is important to notice that SUGAR was not sensitive enough to be able to
 detect the neutron flux which we predicted in Sec.s  \ref{PD} and \ref{neu_pp} to come from Sgr A East. Therefore, the lack of signal in the direction of the GC in the SUGAR data set does not 
 contradict the hypothesis raised in this paper.  
 
 The normalisation of the relativistic particle spectrum in Sgr A East which was adopted 
in  \cite{Crocker} was based on a combined fit of  the low energy $\gamma$-ray signal
observed by EGRET \cite{EGRET} in the direction of
 the GC and the AGASA/SUGAR anisotropy data. As we explained in Sec.\ref{gammas}, and 
noted also by the authors of \cite{Crocker}, the $\gamma$-ray spectrum of the source responsible 
for  the EGRET excess cannot extend up to  very high  energies. It is, therefore, quite unnatural 
to assume that this source accelerates CRs up to the EeV.   Although in \cite{Crocker} the authors
 also (marginally) consider a  HESS + AGASA/SUGAR  fit, in order this to work  they need to
 adopt a power index $1.97$ for the proton spectrum  which is almost $3\sigma$ lower than the HESS best fit \cite{HESS} (as we showed in Sec.\ref{gammas},  the spectrum of high energy secondary photons is expected to have the same slope as that of primary protons).
Such  a discrepancy may become more serious if new data from HESS would  prolong the GC
$\gamma$-ray spectrum observed so far to even higher energies.

Last but not least,  the contribution of nuclei  photo-disintegration to Sgr A East  neutron emission   was not determined in \cite{Crocker}. We showed, however,  that this process
should provide the dominant, if not the unique,  channel for the production of EeV  neutrons. 
 
 In this paper we did not discuss  the contribution to the synchrotron radio emission and to the
 inverse Compton (IC) low-energy $\gamma$-ray flux from Sgr A East which may be  produced by secondary leptons. The possible relevance of these emissions was studied 
in \cite{FatMel03}.   The  authors of \cite{FatMel03} assumed that the $\gamma$-ray flux 
observed by EGRET \cite{EGRET} is a by-product  of $pp$ inelastic  
scattering in Sgr A East and used  EGRET measurements to normalise the primary
proton flux. On the basis of these assumptions, they estimated  the expected  radio and IC secondary fluxes  showing that these emissions  are  compatible with the observations. 
Since the normalisation  of the relativistic nuclei spectrum that we used in this paper
(which is based on HESS observations) is much lower,  we are confident that secondary emissions
are even harmless in our case.  

We also did not discuss here the possible effect of post-acceleration diffusion  on  the primary nuclei spectrum.  
In principle, since diffusion is an energy dependent process,  deviation from a single 
power-law behaviour may  arise in the nuclei spectrum.  Some possible consequences of high energy proton 
diffusion in the  GC region has been recently discussed in \cite{AhaNer05}. 
Here  we implicitly assumed that above several TeV  (the energy required to produce the TeV 
gamma-rays observed by HESS)  protons spectrum  around Sgr A East coincides with the  acceleration spectrum.

 \section{Conclusions}\label{conclusions}  

In this paper we estimated the flux of neutrons reaching the Earth from Sgr A East under the
educated  hypothesis that nuclei are accelerated beyond the EeV in that SNR.
We showed that neutrons with the required energy ($\sim 1 \EeV$) can be naturally produced by
 the photo-disintegration (PD) of composite nuclei onto the IR background 
radiation present in the GC region and by the $pp$ scattering of UHE protons onto the
dense hydrogen gas surrounding Sgr A East. 
We found the former process to be the dominant.  Amazingly,  the  PD rate of nuclei 
onto the 40 K photon background   peaks  just at the energy  required to produce
EeV neutrons which can  reach the Earth from the GC before decaying.
Due to their largest abundance respect to other composite nuclides, 
 $^4$He nuclei should give the main contribution to the neutron flux.   
While the energy threshold for $pp$ scattering is much lower, due to the low neutron elasticity 
(the fraction of  the primary proton energy going to the neutron) the energy required to 
primary protons in order to produce  EeV neutrons is, in average,  higher than 10 EeV.     
Such a high energy may be hardly reachable even in Sgr A East so that the contribution of 
$pp$ inelastic scattering to the EeV neutron flux may be absent.

The signal produced by the EeV neutrons from Sgr A East should  be easily disentangled from the CR isotropic background thanks to its particular spectrum and its point-like nature.
As we showed in Sec.\ref{PD}, the neutron spectrum should  start to rise  rapidly above 
$few \times 10^{17}~\eV$, to peak just below the EeV,   and become a power law  steeper than the
 CR spectrum at higher energy.  The reader should take in mind, however, that a UV cut-off may 
come-in somewhere above the EeV  as the maximal acceleration energy in Sgr A East cannot
be indefinitely high. Presumably, only one energy bin around the EeV will be interested
by the excess.   
The clearest signature of the neutron flux from Sgr A East should be given, however,  by the
 point-like nature of the excess and by its coincidence with the actual position of this SNR
 in the sky coinciding with that of the GC.  
By assuming that nuclei accelerated in Sgr A East have a power law spectrum 
with exponent $- 2.2$,  as suggested by HESS gamma-ray observations,  and that this
spectrum extend steadily beyond the EeV,  we determined the flux of secondary neutrons 
expected to reach the Earth and  showed that the Pierre AUGER Observatory should be able to detect it.
    
 We conclude by observing that several of the results that we derived in this work for Sgr A East
 should also be applicable to some other SNRs. As we discussed in the Introduction, 
 the most promising neutron/gamma-ray  emitters  as those SNRs interacting with dense 
 molecular clouds. The fast improvements in 
gamma-ray astronomy and in the extensive air shower  reconstruction techniques may  
  allow soon  a  systematic search of galactic EeV-trons.
  
 \section*{Acknowledgements}
 
 We  thank F. Aharonian, P. Blasi, V. Cavasinni,  S. Degl'Innocenti, E. Roulet, S. Shore and  M. Vietri,  
 for valuable discussions and in particular F.A., V.C., E.R., S.S. and M.V.  for reading the draft  
 of this paper providing several useful hints and comments.
 This work would not have been possible without the assistance and support of the Pisa
 ANTARES group.  L.M. thanks G. Usai for help using the \textsc{Pythia}  package.
 
\section*{Appendix}

The $pp$ total cross-section determined by combining  collider and UHECR
($pp_{\rm air}$ scattering)  data \cite{PDG}  is
\begin{equation}
\label{sigma_pp}
\sigma_{pp}(s) = Z_{pp}+B\; \ln^2 \left({s \over s_1}\right)
\end{equation}
where  $Z_{pp} = 35.48 \times 10^{-27}~\cm^2$, $B = 0.310 \times 10^{-27}~
\cm^2$  and  $s_1 =  (5.38 ~\GeV)^2$. 

Clearly, we are interested only in the inelastic $pp$ scattering.
 The inelasticity $K$  is defined by $\sigma^{\rm inel}_{pp} = K  \sigma_{pp}$
From the experimental data of  the  $pp$ and $p{\bar p}$ scattering $K \simeq 4/5$. 
For $s > 10~\GeV^2$ there is no evidence of a  variation of $K$ with energy. 

The neutron emissivity can be written in the form  
\begin{equation}
\label{Qn_pp_app}
Q_n^{pp}(E_n) =
c\; n_H \int_{E_p^{min}}^\infty dE_p   ~n_p(E_p)~{\rm n}_n
\sigma^{\rm inel}_{pp} \frac{dP_n}{dE_n}(E_n, E_p)\;, 
\end{equation} 
where ${\rm n}_n$ is the neutron multiplicity and 
$\displaystyle \frac{dP_n}{dE_n}(E_n, E_p)$ is the normalised probability
distribution to produce a neutron with energy $E_n$. 

 Since neither experimental data  nor  undisputed theoretical calculations are available 
 allowing to determine these quantities  at UHEs, we resort  to  Monte Carlo  simulations.
We used the \textsc{Pythia} generator \cite{PYTHIA} to simulate $pp$ inelastic scattering for 
several values of $E_p$  in the range  $1 -10^3~\TeV$
 under the constraint that the total cross section is that given in (\ref{sigma_pp}).
We found  the value of the neutron multiplicity  to be ${\rm n}_n = 0.24$  almost
independently  on $E_p$. 
The simulation also allowed to determine the neutron energy distribution. Again,  
the result was practically energy independent meaning that differential cross section
 $\displaystyle \frac{d\sigma}{dE_n}$ can be written in a scaling form:
 \begin{equation}
 \label{scaling}
\frac{d\sigma}{dE_n}=\sigma_{inel}\,{\rm n}_n~
\frac{dP_n}{dE_n}(E_n, E_p) =   {\rm n}_n~\frac{\sigma^{\rm inel}_{pp}}{E_n}~h(x)\;,
\end{equation} 
where $x \equiv E_n/E_p$.
The best fit to  the Monte Carlo data gives
\begin{equation}
h(x)  \simeq  0.064\left(1-x\right)^2+0.094x\sqrt{1-x} ~.
\end{equation}
This result is similar to that  found by Drury et al. \cite{Drury} but for the last term,
being  absent in their work,  which allow a better fit of the MC data for $x \rightarrow 1$.

The mean value of  $x$,  the so called neutron {\it elasticity},  is 
\begin{equation}
\langle x  \rangle = \int_0^1  dx ~h(x)  \simeq 0.05~,
\end{equation}
 Before inserting the previous  results in (\ref{Qn_pp_app}) we rewrite the inelastic cross section
in the form 
\begin{equation} 
\sigma_{pp}^{\rm inel}(E_p) =
\sigma_0~\left( 1+D \ln\left(\frac{E_p}{E_0}\right)\right)\;,
\end{equation} 
where we defined $\sigma_0 \equiv  \sigma_{pp}(E_0) \simeq K \times Z_{pp}$, 
and $D   \equiv 2 B/Z_{pp}  = 1.7 \times 10^{-2}$ where $E_0 = 1~\GeV$.
The neutron emissivity can  then be written as 
\begin{equation}
Q_n^{pp}(E_n) = c \; n_H  n_p(E_n) \sigma_0  {\rm n_n} K 
 \left[ I_1(\alpha) + I_2(\alpha)  \ln\left(\frac{E_n}{E_0}\right)  \right]
\end{equation}
where
\begin{equation} 
\label{Ialpha}
\begin{split} &I_1(\alpha) \equiv
 \int_0^1dx\,x^{\alpha-2}h(x)(1-D\ln\left(x\right))\\
 &I_2(\alpha) \equiv D\int_0^1 dx\,x^{\alpha-2}h(x)\;. 
 \end{split}
 \end{equation}
 
 Clearly, a regularization procedure is needed in order to avoid singularities near $x = 0$. Obviously the correct regularization is made when one considers the finite mass of the neutron, which forces the integration domain to be $\displaystyle \left[\frac{m_n}{E_p}, 1\right]$.

\end{document}